\documentclass{svjour3}

\usepackage[T1]{fontenc}

\usepackage{hepnames}
\usepackage{blindtext}
\usepackage{mathrsfs}
\usepackage{graphicx}
\usepackage{hepnames}
\usepackage{amsmath}
\usepackage[colorlinks,citecolor=blue,urlcolor=blue,linkcolor=blue]{hyperref}
\usepackage{textcomp}
\usepackage{physics}

\renewcommand{\PV}{\ensuremath{\mathrm{V}}\xspace}
\newcommand{\ttx}{\ensuremath{\Pqt\Paqt}\xspace}
\newcommand{\pvmiss}{\ensuremath{\vec{p}^{\textrm{miss}}_\textrm{T}}\xspace}

\journalname{Eur. Phys. J. C}

\usepackage[pagewise]{lineno}

\begin{document}

\title{MoMEMta, a modular toolkit for the Matrix Element Method at the LHC}


\author{S\'ebastien Brochet
        \and
        Christophe Delaere
        \and
        Brieuc Fran\c{c}ois
        \and
        Vincent Lema\^itre
        \and
        Alexandre Mertens
        \and
        Alessia Saggio
        \and
        Miguel Vidal~Marono
        \and
        S\'ebastien Wertz
}


\institute{Centre for Cosmology, Particle Physics and Phenomenology (CP3),\\
Universit\'e catholique de Louvain, Chemin du Cyclotron 2,
B-1348 Louvain-la-Neuve, Belgium}

\date{January 2019}


\maketitle

\begin{abstract}
The Matrix Element Method has proven to be a powerful
method to optimally exploit the information available in detector data.
Its widespread use is nevertheless impeded by its complexity and
the associated computing time. 
MoMEMta, a \texttt{C++} software package to compute the integrals 
at the core of the method, provides a versatile implementation of the
Matrix Element Method to both the theory and experiment communities.
Its modular structure covers the needs of experimental analysis workflows 
at the LHC without compromising ease of use on simpler and smaller simulated 
samples used for phenomenological studies. 
With respect to existing tools, MoMEMta improves on usability and flexibility.
In this paper, we present version 1.0 of MoMEMta, together with examples 
illustrating the wide range of applications at the LHC accessible for the first 
time with a single tool.
\end{abstract}

\section{Introduction}
\label{sec:intro}

The discovery of the Higgs boson by the ATLAS and CMS experiments in
2012~\cite{HiggsAtlas,HiggsCms} opened a new era in particle
physics. More than just a new particle, a new set of interactions needs
 to be characterised.  The LHC physics program therefore includes
precision measurements of standard model (SM) processes (in particular
in the top-quark and Higgs sectors) and the search for rare production mechanisms or rare decay channels.
The absence so far of any obvious sign of physics beyond the SM further
increases the need to look in places where the backgrounds are large
and the effect of new physics subtle.

In all these studies, it is of the uttermost importance to fully
exploit the potential of the large data set collected.  For most of
the analyses performed in high energy physics (HEP), obtaining an optimal
result implies the treatment of multiple correlated quantities in a
multivariate setting. The most popular methods for multivariate analysis
in HEP are machine learning techniques, such as boosted decision trees and
neural networks. These approaches require large training data sets
(usually obtained by Monte Carlo techniques) in order to learn the
 structure of the data.  On the contrary, the Matrix Element
Method (MEM) uses directly our theoretical knowledge of a process to
assign to each event a probability that measures the compatibility of
experimental data with a given hypothesis.  There is no training,
since the underlying Lagrangian, from which the matrix element of the
partonic process is derived, is known.

The MEM, originally designed at the Tevatron experiments D\O\ and
CDF for top quark mass measurements in $\Ptop\APtop$ production~\cite{Dalitz:1998zn,Aaltonen:2008bg,Aaltonen:2008bd,Aaltonen:2011di,Collaboration:2012hz,Abazov:2015spa,D0:2016ull}, is nowadays a common technique in
particle physics. Recent examples of its use at the LHC are searches for $\Ptop\APtop\PH$~\cite{Aad:2015gra,Aaboud:2017rss,Khachatryan:2015ila,Sirunyan:2018mvw,Sirunyan:2018ygk,Aaboud:2017vzb,Aaboud:2017jvq,Sirunyan:2018shy}
and single top quark production~\cite{Aad:2015upn}, and a measurement of spin correlations in $\Ptop\APtop$ production~\cite{Khachatryan:2015tzo}.
Nevertheless, while it can be used for a wide variety of studies, the practical
application of the MEM has been impeded by its complexity and by the
associated computing time. In order to evaluate the probability under
a given theoretical hypothesis of a given experimental event, a
difficult convolution of the theoretical information on the hard
scattering (i.e. the matrix element squared) with the experimentally
available information on the final state (encoded in the so-called
transfer functions) has to be performed.  The corresponding integrand
varies by several orders of magnitudes in different regions of the
phase space, which requires the use of adaptive numerical integration
techniques together with a smart choice of integration variables.
A general algorithm has been proposed in Ref.~\cite{Madweight}, which
involves optimised phase-space mappings designed to remove as much as possible the peaks in the integrand.  
However, the corresponding implementation
(\textsc{MadWeight}) is not supported anymore and suffers from a lack of
flexibility that prevents---or significantly limits---its use in large
scale analyses of LHC data by the collaborations, 
and does not allow the user to implement simplifying assumptions.

In this paper, we present MoMEMta, a modular \texttt{C++} software package to
compute the convolution integrals at the core of the method. Its modular structure
covers the needs of experimental analysis workflows at the LHC without
compromising the ease of use on simpler and smaller simulated samples
used for phenomenological studies.  It relies on the same approach as
\textsc{MadWeight} to address the parameterisation of the phase space but 
leaves more freedom to the user. Since it follows the same approach, 
MoMEMta's performance in terms of accuracy and CPU time is similar to 
that of \textsc{MadWeight}. But contrarily to its predecessor, 
it adapts to any process and can be fitted to any 
\texttt{C++} or \texttt{Python} analysis workflow.
Modularity and flexibility also open the door to specific optimisations
either when designing the integration structure, or when choosing the integration engine.
It is also possible to provide a custom (optimised) matrix element when appropriate, 
without loosing all the advantages of the block decomposition described in Sec.~\ref{sec:impl}.

In the following, we will first briefly review the MEM, with an emphasis on
the assumptions made in MoMEMta, before presenting shortly the
philosophy of the implementation. We will then concentrate on a few
concrete use cases that illustrate the variety of problems that can be
tackled using MoMEMta, and how the modularity can best be exploited to
adapt to these problems.

\section{The matrix element method}
\label{sec:MEM}

The MEM is a technique to calculate
the conditional probability density $P(x | \alpha)$ to observe an experimental event
$x$, given a specific theoretical hypothesis $\alpha$.
Details about the method can be found for example
in Ref.~\cite{Fiedler:2010sg}. We will here concentrate on the main
aspects.

The likelihood for a partonic final state $y$ to be produced in the
hard-scattering process is proportional to the differential cross
section ${\rm d}\sigma_\alpha$ of the corresponding process, given by

\begin{eqnarray}
  \label{eq:dsigmaP} {\rm d}\sigma_\alpha(q_1,q_2,y) = \frac{(2
\pi)^{4}\!\left| \mathscr{M}_\alpha\left(q_1,q_2,y\right)
\right|^{2}}{ q_1 q_2 s} {\rm d}\Phi(y)\ ,
\end{eqnarray}
where $q_1$ and $q_2$ stand for the initial state
parton momentum fractions, $s$ stands for the hadronic centre-of-mass energy
and $y$ stands for the kinematics of the final state.

The central element in that expression is the squared matrix element for 
process $\alpha$, denoted $|\mathscr{M}_\alpha\left(q_1,q_2,y\right)|^2$, where
the summation over spin and colour states is understood.  It can
be obtained either analytically or numerically through packages like
\textsc{MG5\_aMC@NLO}~\cite{madgraph5} or MCFM~\cite{MCFM}.  Because of the intrinsic
theoretical difficulty to identify final state particles with partons at
NLO, the leading order matrix element is used in most applications.
The n-body phase space ${\rm d}\Phi(y)$ must also be considered
in the calculation, as it plays an important role in any change
of variable needed for the integration of the differential cross
section.

To obtain the differential cross section ${\rm d}\sigma_\alpha(y)$ in hadron collisions, (\ref{eq:dsigmaP}) is convoluted with the parton density functions (PDF) and summed over all possible flavour compositions of the colliding partons,

\begin{equation}
  \label{eq:dsigmaPpp} {\rm d}\sigma_\alpha(y) = \int\limits_{q_1,
q_2} \! \sum_{a_1, a_2} {\rm d}q_1 \, {\rm d}q_2 \, f_{a_1}(q_{1}) \, f_{a_2}(q_{2})\
{\rm d}\sigma_\alpha(q_1,q_2,y),
\end{equation}
where $f_{a_1}(q_{1})$ and $f_{a_2}(q_{2})$ are the PDFs for a given flavour $a_i$ and momentum fraction $q_i$.

The evolution of the parton-level configuration $y$ into a
reconstructed event $x$ in the detector is modelled by a transfer
function $T(x|y)$, normalised as a probability density over $x$,
that describes how the partonic final state $y$ is
reconstructed as $x$ in the detector. This includes the effects
from the parton shower, hadronisation, and the limited detector resolution.
The efficiency $\epsilon(y)$, i.e. the probability to reconstruct 
and select a specific partonic configuration $y$, also needs to be taken into account.
This includes geometrical acceptance effects.

The transfer function and efficiency are assumed to factorise into contributions from
each measured final-state particle, and each of these contributions are 
often assumed to further factorise into simple direction- and momentum-dependent terms.
For most applications, it is realistic to then assume that
particle directions are perfectly reconstructed. The transfer function in that
case is a Dirac delta function on the angular variables, and takes a non-trivial
form only for the energy (or transverse momentum) degree of freedom.
All these assumptions may not be valid
in cases where objects, especially jets, are close to each other or
reconstructed together (e.g. in boosted topologies), which will then
result in a less accurate result.

Once the constraints related to the kinematics of observed particles have been taken into account, there may remain unobserved degrees of freedom, such as those pertaining to neutrinos (or any other invisible particles) or to unreconstructed objects outside of detector acceptance, as well as to the initial-state partons.
Some of these degrees of freedom can be removed by enforcing the conversation of total 4-momentum in the initial and final states; the remaining ones need to be marginalised, resulting in a potentially large volume of phase space over which to integrate.
Additional constraints may then be used to reduce this volume, such as mass constraints from intermediate, narrow resonances in the considered process, assumed to be on their mass shell, or the experimentally measured total transverse momentum $\pvmiss$ of the missing particles in the event.
Note that while a resolution function can be built on $\pvmiss$, it is not accurate to trivially factorise the transfer function on the two components of $\pvmiss$ from the terms relative to visible particles in the final state, since the experimental error in measuring $\pvmiss$ is correlated with the error made in measuring all the other particles in the event.

Aspects to be considered in the transfer function and efficiency are the measurement
of the momentum of a particle as well as its (mis-)identification.
This latter point might be relevant for b quarks and $\tau$ leptons,
and allows in principle to combine different partonic final state hypotheses for the same event.

After convolution, the full expression reads
\begin{subequations}\begin{align}
  \label{eq:mem2}
  P(x|\alpha)   & = \frac{1}{\sigma_{\alpha}^{\text{vis}}} \int \!
                        {\rm d}\sigma_\alpha(y) \, T(x|y) \, \epsilon(y) \\ 
                & = \frac{1}{\sigma_{\alpha}^{\text{vis}}} 
                        \int \limits_{q_1,q_2} \! \sum_{a_1, a_2}\int \limits_{y} \!
                        {\rm d}\Phi(y) \, {\rm d}q_{1} {\rm d}q_{2} 
                        f_{a_1}(q_{1}) \, f_{a_2}(q_{2}) \, 
                        |\mathcal{M}_{\alpha}(q_1,q_2,y)|^{2} \, T(x|y) \, \epsilon(y),
  \label{eq:mem1}
\end{align}\end{subequations}
where $\sigma_{\alpha}^{\text{vis}}$ is a normalisation factor that ensures $P(x | \alpha)$ is a
probability density over $x$.
While that factor can be computed by explicitly integrating $P(x|\alpha)$ over the phase space of reconstructed events $x$, it is often more practical to estimate it using a sample of events simulated under hypothesis $\alpha$, in which case $\sigma_{\alpha}^{\text{vis}} = \sigma_{\alpha} \cdot \expval{\epsilon}_{\alpha}$, where $\sigma_{\alpha} = \int {\rm d}\sigma_\alpha(y)$, and $\expval{\epsilon}_{\alpha}$ is the average reconstruction and selection efficiency of the simulated events.
Finally, one has also to take into account the fact that
some of the particles measured in the detector cannot be assigned
unambiguously to specific final-state partons. Generally, all possible combinations
have then to be considered and the resulting values for $P(x|\alpha)$ averaged.

The information contained in (\ref{eq:mem1}) can be exploited in
different ways, from the extraction of the most probable value of 
theory parameters through a likelihood maximisation method (see e.g.~\cite{FerreiradeLima:2017iwx}), 
for which the dependence of the normalisation constant $\sigma_{\alpha}^{\text{vis}}$ on the considered hypothesis has to be properly taken into account,
to the bare use of the integral result without normalisation 
$\sigma_{\alpha}^{\text{vis}}$, referred to in the
literature as matrix element weight, $W(x|\alpha)$.

The integral defined in (\ref{eq:mem1}) is typically a small
number that varies over several orders of magnitudes from event to
event.  It is therefore common to use instead the \emph{event information} defined by $I_{\alpha} \equiv -\log P(x|\alpha)$.  When
computed from the weight instead of the probability, the information
is only modified by an additive constant, with no consequence in 
many applications. We will denote this quantity $I'_\alpha \equiv -\log W(x|\alpha)$.

In the limit where all the quantities and functions in
(\ref{eq:mem1}) are known with perfect accuracy, 
$P(x|\alpha)$ is a likelihood.  By the Neyman--Pearson
lemma, the ratio between the likelihoods obtained under two different
hypotheses $\alpha$ and $\alpha'$ is the most powerful test statistic
to discriminate one from the other~\cite{NeymanPearson}.  Hence, if it
can be implemented, the MEM should provide optimal experimental sensitivity.
In practice, we are limited by the use of leading-order
matrix elements, or by assumptions made in constructing the transfer function and efficiency term. 
The quantity (\ref{eq:mem1}) is therefore not a true
likelihood, and the Neyman--Pearson lemma does not strictly apply.  For
discrimination purposes, it is then common to use the event
information as input of another multivariate method (typically a
boosted decision tree or a neural network).

\section{Implementation}
\label{sec:impl}

The MEM is used in HEP by both theoretical and experimental communities with different purposes and levels of 
complexity ranging from the evaluation of a matrix element on a reconstructed event to the precise evaluation of model parameters (e.g. the top quark mass) through the use of the properly normalised likelihood derived from (\ref{eq:mem1}).
Note that in the former case no integration process is required and effects related to parton showering, hadronisation, and finite detector resolution are explicitly neglected.

In order to adapt to these very different use cases, a novel modular design has been adopted for MoMEMta. 
The core library is written in \texttt{C++} and provides modules for various purposes:
to represent and evaluate the matrix element and parton density functions,
to represent and evaluate transfer functions,
to perform changes of variables,
to handle the combinatorics of the final state, etc.
That way, every term of (\ref{eq:mem1}) is treated as a module that can be configured by the user. 
Weights are computed for a given process by calling and linking the proper set of modules in a 
configuration file written in the \texttt{Lua} scripting language~\cite{lua}. 
Thanks to this modularity, the user is free to substitute any module provided with a custom implementation
without loosing the benefits of other parts of the tool. 
The resulting object can be called from any \texttt{C++} or \texttt{Python} code, which means that it seamlessly 
integrates into the complex analysis environment of the large experimental collaborations but can also be used within 
small programs reading events from files in any format (e.g. a custom text file, or a file in the 
\textsc{Root}~\cite{ROOT}, \textsc{HepMC}~\cite{hepmc}, \textsc{Lhco}~\cite{lhco} or \textsc{StdHEP}~\cite{stdhep} 
format).
In case the modules shipped out-of-the-box are not sufficient for a particular application, it is straightforward for the user to extend MoMEMta's functionalities by adding new modules handling a specific task, still profiting from the existing infrastructure provided by the tool.

The computation of the weights requires, in most cases, the
evaluation of multidimensional integrals via adaptive Monte Carlo
techniques. The efficiency in computing these integrals depends on the
parameterisation of the phase-space measure used in the integration. In
order to map in an efficient way all the structures in the integrand,
MoMEMta follows the philosophy introduced by
\textsc{MadWeight}~\cite{Madweight}. In this approach, starting from a standard
parameterisation, the phase-space measure is optimised by using a
finite number of \emph{analytic} transformations over subsets of the integration variables,
called ``blocks''. 
A list of the blocks available in MoMEMta along with
the addressed event topologies, and the integration
variables removed and introduced by the changes of variables, is shown in
Tabs.~\ref{table:blocks} and~\ref{table:secondblocks}. 
For consistency, we have adopted the same terminology as in Ref.~\cite{Madweight}.
Since the considered transformations are nonlinear, specifying the value of the new integration variables typically yields several solutions for the canonical variables.
The matrix element, transfer function, efficiencies and PDFs all need to be evaluated on each of those solutions, and summed to define the final integrand.
The first table lists ``Main Blocks'', i.e. changes of variables that
allow to integrate out the four-dimensional Dirac delta function present in the phase-space
density term ${\rm d}\Phi(y)$, that enforces conservation of momentum between
the initial and final states.
The second table lists ``Secondary Blocks'', i.e. simple changes of variables
that do not remove any degree of freedom.
Main and secondary blocks are implemented as dedicated MoMEMta modules, 
which can be chained to perform the change of integration
variables that is best suited for the problem at hand. 
These modules also take care of computing the jacobian factors required by the changes of integration variables, to be multiplied with the considered integrand.
Examples of using these blocks are given in Sec.~\ref{sec:usecases}.

\begin{table}[!htpb]
\caption{
Set of MoMEMta Main Blocks. Each block performs a specific change 
of integration variables, and removes four degrees of freedom by
enforcing momentum conservation between the initial and final states.
The third and fourth columns show
the integration variables respectively removed and introduced in each
block definition. Block G is introduced in this work, in addition
to those originally defined in \textsc{MadWeight}. 
We denote by $q_i$ the Bjorken fractions of the initial-state partons, and by $p_i$ the 4-momentum of a final-state particle, parameterised in polar coordinates by $|p_i|$, $\theta_i$ and $\phi_i$.
An off-shell particle decaying to a set of on-shell final-state particles is written $s_{i \dots j} ( \to p_i \dots p_j )$, and the quantity  $s_{i \dots j}$ is defined as $(p_i + \dots + p_j)^2$.
In block E, $y$ denotes the rapidity of the total partonic system.
Removing a particle, $p_i$, means removing all three degrees of freedom associated with that particle.
Variables that are not explicitly removed are understood to remain present as in the standard polar phase-space parameterisation.
Similarly, additional final-state particles not mentioned in the block topology are allowed. Secondary blocks may be used to transform the corresponding phase-space variables.
}
\label{table:blocks}
\vspace*{\medskipamount}
\centering
\small
\begin{tabular}{l c c c}
\hline
\hline
Main & Topology &  Removes\dots & For\\
block & & & \\
\hline
A & $(q_1,q_2) \to p_1 + p_2$ &$q_1$, $q_2$, $|p_1|$, $|p_2|$ &  \\
B & $(q_1,q_2) \to s_{12} (\to \, p_1 + p_2) $ &$q_1$, $q_2$, $p_1$  & $s_{12}$  \\
C & $(q_1,q_2) \to s_{123} \to p_3 + s_{12} (\to \, p_1 + p_2) $ & $q_1$, $q_2$, $p_1$,  $|p_3|$& $s_{12}$, $s_{123}$ \\
D & $(q_1,q_2) \to s_{134} (\to p_4 + s_{13} (\to \, p_1 + p_3)) +$ & $q_1$,$q_2$, $p_1$, $p_2$& $s_{13}$, $s_{134}$, $s_{25}$, $s_{256}$  \\
  & $s_{256} (\to p_6 + s_{25} (\to \, p_2 + p_5))$ & & \\
E & $(q_1,q_2) \to (s_{1234},y)\to s_{13} (\to \, p_1 + p_3) + $ & $q_1$, $q_2$, $p_1$, $p_2$ & $s_{1234}$, $y$, $s_{13}$, $s_{24}$  \\
  & $s_{24} (\to \, p_2 + p_4)$ & &  \\
F & $(q_1,q_2) \to s_{13} (\to \, p_1 + p_3) + s_{24} (\to \, p_2 + p_4)$ &$p_1$, $p_2$ & $q_1$, $q_2$, $s_{13}$, $s_{24}$ \\
G & $(q_1,q_2) \to s_{12} (\to p_1 + p_2) + s_{34} (\to p_3 + p_4)$ & $q_1$, $q_2$, $|p_1|$, $|p_2|$, $|p_3|$, $|p_4|$  & $s_{12}$, $s_{34}$ \\
\hline
\hline
\end{tabular}
\end{table}

\begin{table}[!htpb]
\caption{Set of MoMEMta Secondary Blocks. Each block performs a specific change 
of integration variables, acting exclusively on final-state particles. The third
and fourth columns show the integration variables respectively removed and
introduced by each change of variables.
Removing a particle, $p_i$, means removing all three degrees of freedom associated with that particle.
Variables that are not explicitly removed are understood to remain present as in the standard polar phase-space parameterisation.}
\label{table:secondblocks} \vspace*{\medskipamount}
\centering
\begin{tabular}{l c c  c } 
\hline \hline 
Secondary & Topology & Removes\dots & For \\
block & & & \\
\hline 
A & $s_{1234} \to (s_{123} \to s_{12} (\to \, p_1 + p_2) + p_3 ) +
p_4$ & $p_1$ & $s_{1234}$, $s_{123}$, $s_{12}$ \\
B & $s_{123} \to s_{12} (\to \, p_1 + p_2) + p_3$ & $|p_1|$, $\theta_1$ &
$s_{12}$, $s_{123}$ \\
C/D & $s_{12} \to \, p_1 + p_2 $ & $|p_1|$ & $s_{12}$ \\
E & $s_{123} \to s_{12} (\to \, p_1 + p_2 ) + p_3$ & $|p_1|$, $|p_2|$ &
$s_{12}$, $s_{123}$ \\
\hline \hline
\end{tabular}
\end{table}

In order to efficiently handle the potentially large combinatorial ambiguity in the assignment between reconstructed final-state objects and partons in the matrix element, we have included a dedicated module to average over all permutations between a given set of particles.
This module requires an additional dimension for the integrated phase space, so that the associated variable governs which assignment should be used for the computation of the integrand, and the resulting integral corresponds to the average weight over the considered permutations.
However, compared to a naive averaging of the possible assignments, this scheme allows adaptive integration algorithms to concentrate on those yielding the largest contribution to the final result.
For a fixed number of evaluations of the computationally expensive parts of the integrand, such as the squared matrix element, the precision on the result is thereby increased.

MoMEMta ships with matrix elements for a few processes, but any leading-order process handled by 
\textsc{MG5\_aMC@NLO}~\cite{madgraph5} can be added using a Matrix Element Exporter plugin provided~\cite{MoMEMta-MaGMEE}.
Native support for other matrix element generators is planned for future releases, but the modular structure already 
enables the user to wrap any \texttt{C++} code that computes a matrix element to be used with MoMEMta.
This novel feature can potentially speed up the computation by a substantial amount, since the evaluation of the matrix element largely 
dominates the computation time.

Parton density functions are obtained from \textsc{Lhapdf6}~\cite{Buckley:2014ana} and the integration is done using the 
\textsc{Cuba}  library~\cite{Hahn:2004fe}, that offers a choice of four independent routines for multidimensional 
numerical integration: Vegas~\cite{vegas}, Suave~\cite{Hahn:2004fe}, Divonne~\cite{divonne}, and 
Cuhre~\cite{cuhre1,cuhre2}.

The MoMEMta implementation~\cite{MoMEMtaImpl} is publicly available, is licensed under the GLPv3, and comes together with 
an online documentation~\cite{momemta.github.io} and tutorials~\cite{tutorials}.

\section{MEM use cases}
\label{sec:usecases}

One common application of the MEM is parameter estimation, through which one can extract a parameter of interest by means of likelihood maximisation.
Nowadays, applications of the MEM in high-energy physics are most often restricted to computing weights $W(x|\alpha)$ under several hypotheses,
to discriminate a signal from one or several backgrounds. MoMEMta fits the needs for either purpose, since it is specifically designed to efficiently compute integrals as defined in (\ref{eq:mem1}).
In complex situations with several reconstructed
objects and unconstrained degrees of freedom, there is no unique 
solution to the problem of efficiently and precisely computing $W(x|\alpha)$, 
and the user has to play an active role in
defining how to evaluate $W(x|\alpha)$ and at which accuracy.

In this section we describe a few use cases of the MEM for signal
extraction in LHC analyses. The examples illustrate various levels of
complexity, from the simplest case with a precisely reconstructed
final state for which no integration is needed, to complex final
states including six reconstructed and two unobserved objects.
The \texttt{Lua} configurations for each of the examples can be found together with the MoMEMta tutorials~\cite{tutorials}.
For all the examples, simulated events are generated using
\textsc{MG5\_aMC@NLO}~\cite{madgraph5},
\textsc{Pythia}~\cite{pythia82} and \textsc{Delphes}~\cite{delphes3}.

The computation times vary by several orders of magnitude among the different use cases, and strongly depend on the choice of parameters governing the integration procedure.
Indicative performance figures are given in Sec.~\ref{sec:perf}.

\subsection{Discovery and characterisation of the Higgs boson}
\label{sec:mela}

The MEM was instrumental for the CMS collaboration in the discovery 
of the Higgs boson
in the $\rm{H} \to \PZ\PZ^* \to 4 \Pl$ channel~\cite{HiggsCms}.
Likewise, the characterisation by ATLAS~\cite{Aad:2014eva,Aad:2015mxa}
and CMS~\cite{Sirunyan:2017exp,Sirunyan:2017tqd} of the discovered
resonance in terms of coupling structure, spin and parity, has relied
on matrix-element techniques as suggested
in Refs.~\cite{Bolognesi:2012mm,Anderson:2013afp}.  In this channel, all
final-state particles can be detected and there are no unobserved
degrees of freedom over which to integrate.  Given the good experimental
resolution on muon and electron direction and momentum, it is
reasonable to approximate the transfer function in (\ref{eq:mem1}) by
$T(x,y) = \delta(x,y)$.  Hence, the integral reduces to a simple
evaluation of the matrix element squared and the PDFs using the
measured momenta in the event.  In this framework, dubbed matrix
element likelihood analysis (MELA), it is
straightforward to build a discriminating variable between the signal
and the $\Pq\Paq \to \PZ\PZ/\PZ\Pgg^* \to 4 \Pl$ background by
considering the matrix elements of these two hypotheses:

\begin{equation}
    \label{eq:mela_bkg} \mathcal{D}_{\text{bkg}}(x) = \left( 1 +
\frac{P(x|\text{bkg})}{P(x|\text{sig})} \right)^{-1}. 
\end{equation}

Similar variables can be constructed to discriminate, for instance, a
SM Higgs boson ($J^P=0^+$) from a resonance of the same mass but
different spin and/or opposite parity:

\begin{equation}
    \label{eq:mela_0m} \mathcal{D}_{J^P}(x) = \left( 1 +
\frac{P(x|J^P)}{P(x|0^+)} \right)^{-1}.
\end{equation}

\begin{figure}[!htbp]
  \centering
    \includegraphics[width=0.45\textwidth]{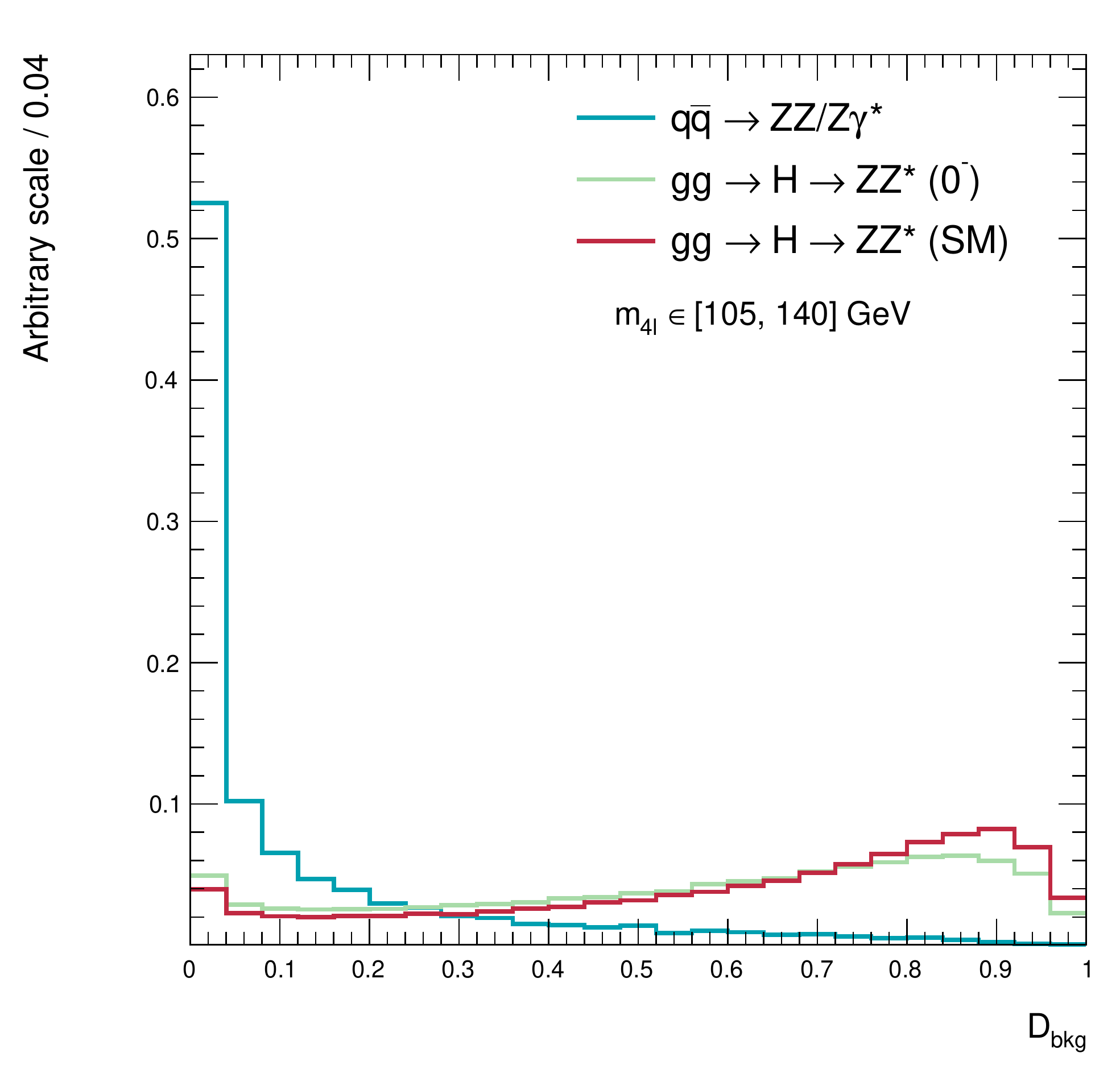}
    \includegraphics[width=0.45\textwidth]{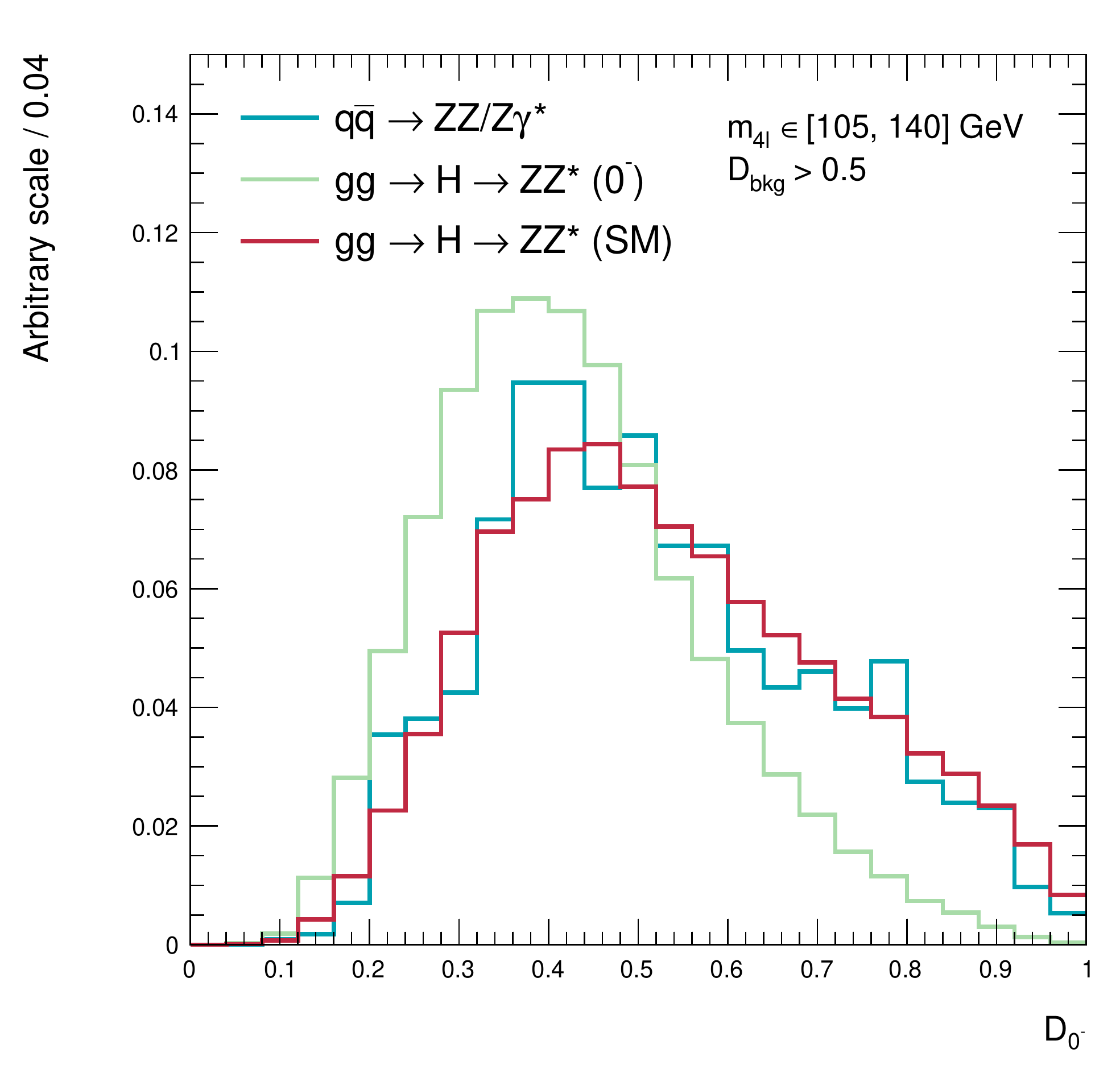}
    \caption{Distribution of the $\mathcal{D}_{\text{bkg}}$
      (left) and $\mathcal{D}_{0^-}$ (right) variables for the $\rm{gg} \to
      \rm{H} \to \rm{ZZ}^* \to 4 \rm{\mu}$ and $\rm{q}\bar{\rm{q}} \to
      \rm{ZZ}/\rm{Z\gamma}^* \to 4 \rm{\mu}$ processes, where the resonance
      $H$ is taken to be the SM Higgs or a pseudoscalar of the same mass
      ($0^-$). For the right-hand figure, we require
      $\mathcal{D}_{\text{bkg}} > 0.5$.
      Note that both the background and the SM Higgs processes have similar distributions of the $\mathcal{D}_{0^-}$ discriminant.
      All distributions are normalised to unit area.}
    \label{fig:mela}
\end{figure}

Although MoMEMta was designed to handle more complex final states, its
flexibility allows the user to easily implement a MELA-like analysis.
To illustrate this fact, we have simulated events for the $\rm{gg} \to
\rm{H} \to \PZ\PZ^* \to 4 \rm{\mu}$ and $\Pq\Paq \to
\PZ\PZ/\PZ\Pgg^* \to 4 \rm{\mu}$ processes.
The SM Higgs sample, as well as the production and decay of a
resonance of spin/parity $J^P=0^-$ were generated using the Higgs
characterisation framework~\cite{higgs_character}.  With MoMEMta's
plugin for \textsc{MG5\_aMC@NLO}, the corresponding matrix elements can be exported in a format suitable for MoMEMta.
The configuration of MoMEMta in this use case only requires a single
module, which returns the product of the matrix element and the PDFs
evaluated on a given event.  The phase-space density term present in
(\ref{eq:mem1}) does not need to be included, since it cancels in the
ratios in (\ref{eq:mela_bkg}) and (\ref{eq:mela_0m}).  With
$P(x|\text{bkg})$, $P(x|\text{sig})$ and $P(x|0^-)$ computed by
MoMEMta, the discriminant variables $\mathcal{D}_{\text{bkg}}$ and
$\mathcal{D}_{0^-}$ can be built.  The distributions of
these variables, for the different processes considered, are shown on
Fig.~\ref{fig:mela} after an event selection closely following the
analysis in Ref.~\cite{Sirunyan:2017tqd}.  The discrimination power between
the competing hypotheses is comparable to what is obtained
in Refs.~\cite{Sirunyan:2017exp,Sirunyan:2017tqd}.

\subsection[Charge identification in tW production]{Charge identification in $\Pqt\PW$ production}
\label{sec:tW}

The MEM has been extensively used in the study of single top quark production processes at the Tevatron~\cite{Aaltonen:2009jj,Abazov:2009ii}, and was instrumental in the most sensitive search for $s$-channel single top production at the LHC~\cite{Aad:2015upn}.
Incidentally, single top and $\PW$ boson associated production ($\Pqt\PW$) provides a good opportunity to showcase MoMEMta's abilities. 
This process features three propagators in the matrix element and, in
the case where both the top quark and $\PW$ boson decay leptonically (dilepton channel), missing information due to the presence of two neutrinos in the final state. 

We consider the charge-conjugate processes, $\Pqt\PWminus$ and $\Paqt\PWplus$, which yield the same visible final state and have practically the same rate at the LHC.
It has been suggested to measure the CKM matrix element $|V_{\Pqt\Pqd}|$ at the LHC using the charge asymmetry between these two processes~\cite{Alvarez:2017ybk}. 
This requires the ability to efficiently disentangle them, a task made difficult by the system not being entirely reconstructible in the dilepton channel.

We thus suggest to construct a MEM-based observable as:

\begin{equation}
    \mathcal{D}_{\pm}(x) = \frac{W(x| \Paqt\PWplus) - W(x| \Pqt\PWminus)}{W(x| \Paqt\PWplus) + W(x| \Pqt\PWminus)}.
    \label{eq:D_pm}
\end{equation}

The charge asymmetry can then be defined by counting the number of events for which either $\mathcal{D}_{\pm}(x) < 0$ or $\mathcal{D}_{\pm}(x) > 0$.

\begin{figure}[!htbp]
  \centering
    \includegraphics[width=0.45\textwidth]{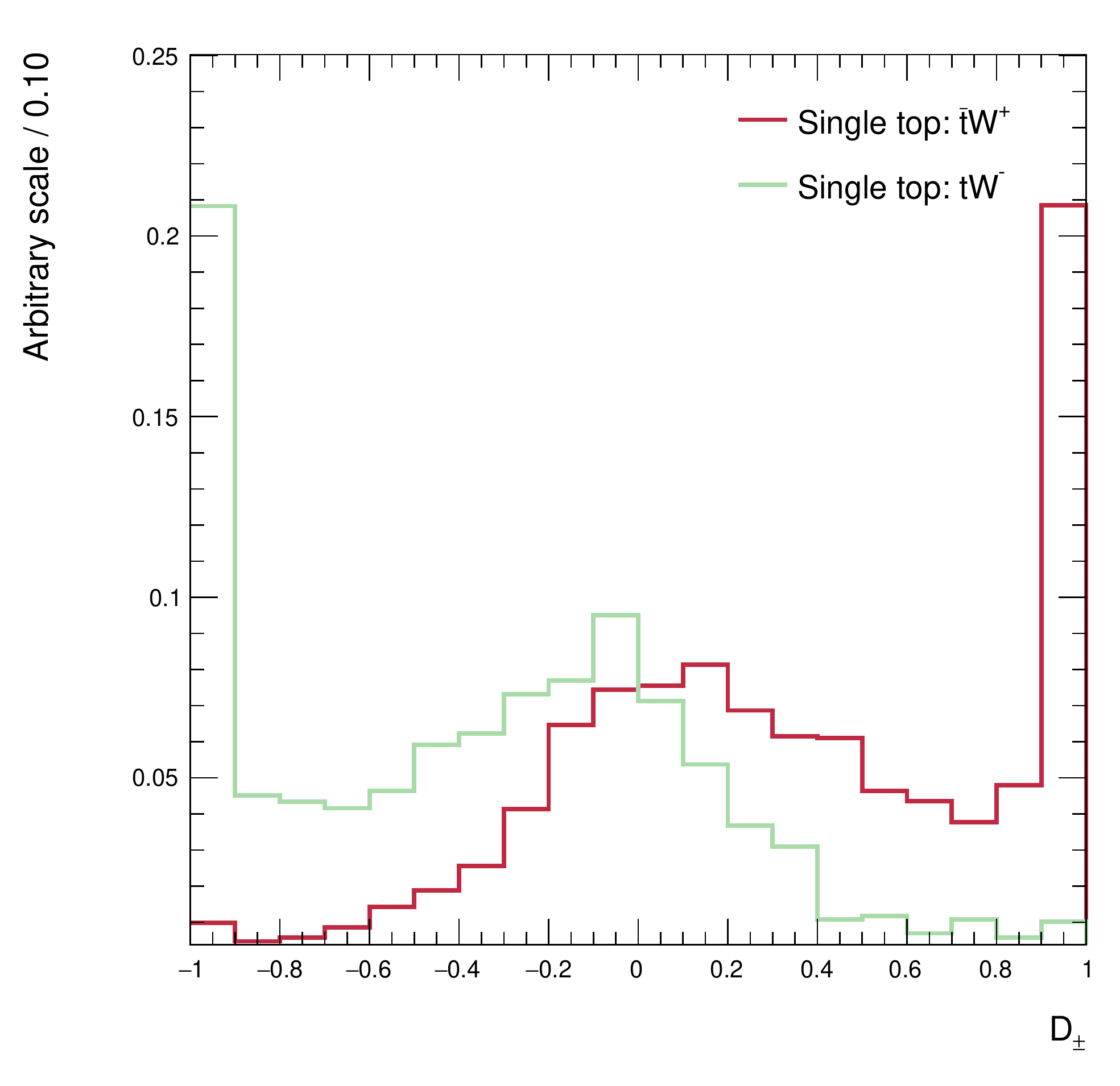}
    \caption{Normalised distribution of the $\mathcal{D}_{\pm}$ asymmetry observable, as defined in (\ref{eq:D_pm}), for the charge-conjugate processes $\Pqt\PWminus$ and $\Paqt\PWplus$. Backgrounds such as $\ttx$ production are expected to be distributed symmetrically around zero.
    }
    \label{fig:tw_charge}
\end{figure}

Computing the weights $W(x| \Paqt\PWplus)$ and $W(x| \Pqt\PWminus)$ requires a careful consideration of the constraints and degrees of freedom at hand.
We start by assuming that the directions of all ``visible'' objects ($\Pqb$ quark, leptons) are perfectly reconstructed, so that the transfer function reduces to factorised parameterisations of the resolution on their energies.
Thus, 11 degrees of freedom are present in the system: the longitudinal momentum of the initial-state partons (2), the energies of the visible particles in the final state (3), as well as the directions and energies of the two neutrinos (6). 
Enforcing conservation of energy and momentum between the initial and final state will remove four of these, so that we end up with seven dimensions over which to integrate.

The numerical integration will be most efficient if the integration variables are mapped to the six peaks in the integrand generated by the top quark and $\PW$ boson propagators and by the transfer functions on the energies of the visible particles; the remaining degree of freedom can be chosen freely.
This can be easily achieved in MoMEMta by pairing the ``Secondary Block B'' with the ``Main Block B''.

We identify the decay chain $\Pqt \to \PW_{\Pqt} (\to \Pnu_{\Pqt} + \Pl_{\Pqt}) + \Pqb $ with the notation $s_{123} \to s_{12} (\to p_1 + p_2) + p_3$ for the chosen secondary block in Tab.~\ref{table:secondblocks}.
The secondary block does not remove any degree of freedom and simply exchanges the energy and polar direction of $\Pnu_{\Pqt}$ for the squared invariant masses of $\Pqt$ and $\PW_{\Pqt}$, which are taken as integration variables.
Since the energies of $\Pl_{\Pqt}$ and $\Pqb$ are associated with peaks in the transfer function, they should be retained as integration variables, which is straightforward since the chosen block does not affect these quantities.
There remains a free variable over which to integrate, the azimuthal direction of $\Pnu_{\Pqt}$, which is not directly associated with any peak in the integrand and can be kept as is.
Given fixed values for $s_{123}$, $s_{12}$ and $\phi_1$, as well as $p_2$ and $p_3$, the block solves the following nonlinear system:
\begin{align}
    s_{12} &= (p_1 + p_2)^2 \\
    s_{123} &= (p_1 + p_2 + p_3)^2 \\
    p_1^2 &= m_1^2 = 0.
\end{align}
Thus, the block returns up to two solutions for the full kinematics of $p_1$, as well as the jacobian factor associated with this change of variables. 
Each of these solutions is used as input for the rest of the computation, described below.

The remaining phase-space variables to be considered are related to the other $\PW(\to \Pnu + \Pl)$ decay as well as to the initial-state partons. 
This system can be identified with the topology $(q_1,q_2) \to s'_{12} (\to p'_1 + p'_2)$ in Tab.~\ref{table:blocks}.
Again, the energy of the charged lepton $\Pl$ is chosen as integration variable.
The main block removes four degrees of freedom by enforcing the conservation of total 4-momentum in the initial and final states, which results in a single variable left to integrate over, chosen as the squared invariant mass $s'_{12}$ of the $\PW$ boson, conveniently aligned with the last peak in the integrand.
The system solved by the main block is:
\begin{align}
    s'_{12} &= (p'_1 + p'_2)^2 \\
    (p'_1)^2 &= (m'_1)^2 = 0 \\
    p'_{1x} &= \vec{p}^{\textrm{miss}}_x - p_{1x} \\
    p'_{1y} &= \vec{p}^{\textrm{miss}}_y - p_{1y},
\end{align}
where $\vec{p}^{\textrm{miss}}_{x,y}$ are the measured components of the missing transverse momentum along the $x$ and $y$ axes.
The block yields up to two solutions for the neutrino $p'_1$ and evaluates the jacobian factor associated with the integration of a four-dimensional Dirac delta as well as with the change of variables.

Using the solutions obtained for the final-state partonic systems, the kinematics of the initial-state partons can now be computed using a dedicated module.
Since the measured $\vec{p}^{\textrm{miss}}_{\textrm{T}}$ has been used to constrain the $\vec{p}_{\textrm{T}}$ of the pair of neutrinos, the total $\vec{p}_{\textrm{T}}$ of the system is not guaranteed to vanish.
As suggested in Ref.~\cite{Alwall:2010cq}, we thus apply a transverse boost on the final-state system to a frame of reference where its total $\vec{p}_{\textrm{T}}$ is zero, compute the longitudinal components of the two initial-state partons in that frame, and boost the complete system back to the original laboratory frame.
This procedure can be understood as a way to correct the effect of initial-state radiation in the observed event.
Finally, using the (up to) four obtained solutions for the full partonic system, the transfer functions, jacobians, PDFs and squared matrix element are evaluated and the results are summed to define the desired integrand function.
Note that the enhancements in the matrix elements due to the top quark and $\PW$ boson propagators can easily be removed by further well-known transformations applied to $s_{123}$, $s_{12}$ and $s'_{12}$.
These transformations are handled by specialised modules.

The distribution of the $\mathcal{D}_{\pm}$ discriminant is shown on Fig.~\ref{fig:tw_charge}. 
About 75\% of events from either process can be retained on 
each side of $\mathcal{D}_{\pm} = 0$, which by symmetry leads to a
corresponding mistag rate of 25\%. Depending on the analysis needs, the purity can be 
further improved at the cost of efficiency (e.g. we obtain 1.5\% mistag rate for 25\% efficiency).

The strategy adopted above for the phase-space integration is by no means unique. 
We stress that thanks to the modularity of MoMEMta, it is easy for the user to quickly test alternate approaches.
For instance, working in the narrow-width approximation (NWA) is simply achieved by removing the modules handling the integration over the propagator invariant masses and fixing these to chosen pole masses. Thanks to the changes of variables applied, the kinematic constraints in the system are automatically satisfied.
Modifying the assumptions underlying the transfer functions is equally easy, by configuring, adding or removing modules representing the finite resolution on the kinematics of final-state partons.
The user might also choose to enforce that the total $\vec{p}_{\textrm{T}}$ of the partonic system be zero in the laboratory frame, which is achieved by configuring the main block so that the $\vec{p}_{\textrm{T}}$ of the neutrino represented by $p'_1$ balances that of all other final-state particles.

\subsection[ttH production]{$\Pqt\Paqt\PH$ production}
\label{sec:ttH}

One of the most successful uses of the MEM at the LHC can be found in
the searches for $\ttx\PH$ production.  The ATLAS and CMS
collaborations have applied the MEM in final states with $\PH \to
\Pqb\Paqb$~\cite{Aad:2015gra,Aaboud:2017rss,Khachatryan:2015ila,Sirunyan:2018mvw,Sirunyan:2018ygk}, and
multi-lepton final states with either $\PH \to \PV\PV^*$, where
$\PV=\PW$ or $\PZ$, or $\PH \to \Ptau \Ptau$~\cite{Aaboud:2017vzb,Aaboud:2017jvq,Sirunyan:2018shy}.

Here we demonstrate MoMEMta's ability to efficiently handle processes
as complex as $\ttx\PH$, featuring a large final-state multiplicity,
several propagator enhancements in the matrix element, missing
information due to neutrinos, and many possible jet-parton
assignments.  We consider the channel where the Higgs boson decays to
$\Pqb\Paqb$ and both top quarks decay leptonically, for which the main
irreducible background consists of $\ttx \Pqb\Paqb$ associated
production. 
The relevance of the MEM in this channel was first demonstrated in Ref.~\cite{Artoisenet:2013vfa}.
We generate samples for signal and background processes
and select events with two opposite-charge leptons and at least four $\Pqb$-tagged
jets.

Weights are computed with MoMEMta under two hypotheses, $\ttx\PH(\Pqb\Paqb)$ and $\ttx \Pqb\Paqb$.
The strategy adopted to parameterise the phase space is in many ways similar to that described in Sec.~\ref{sec:tW}, and will only be briefly summarised here.
The assumptions related to the transfer function are the same as those considered for the previous example, with the exception of what concerns the energy of the two charged leptons, assumed to be perfectly measured.
For both hypotheses, the energies of the $\Pqb$ quarks coming from the decays of the top quarks are retained as integration variables.
In order to reduce the number of dimensions over which to integrate, we work in the narrow-width approximation (NWA), by which the $\PW$ boson and top quark propagators are approximated by Dirac delta functions.

The momenta of the two unobserved neutrinos (six degrees of freedom) can be fixed using four constraints corresponding to the top quark and W boson invariant masses, as well as by the requirement that their combined transverse momentum equals the observed transverse missing momentum in the event.
This solving strategy can be implemented by applying the change of variable ``Main Block D'' on the standard phase-space parameterisation for the decay products of the top quarks, and fixing the invariants associated with the top quark and W boson propagators ($s_{134}$, $s_{256}$, $s_{13}$ and $s_{25}$ in Tab.~\ref{table:blocks}) to their respective pole masses.
The remaining degrees of freedom are handled differently, depending on the hypothesis:

\begin{itemize}
    \item $\ttx \Pqb\Paqb$: The standard polar phase-space parameterisation for the two extra $\Pqb$ quarks is retained, i.e. we integrate over both their energies.
    \item $\ttx\PH$: We integrate over the energy of one of the $\Pqb$ quarks from the Higgs boson decay. 
    The other quark's energy is fixed by the requirement that the pair's invariant mass be equal to the true Higgs boson mass (NWA). 
    In MoMEMta, this is achieved by applying the transformation of the ``Secondary Block C/D'', and fixing $s_{12}$ to $m^2_{\PH}$.
\end{itemize}

Using the above parameterisation, the peaks in the integrand remain mapped to the integration variables, and the unobserved degrees of freedom due to the two neutrinos in the final state are effectively removed. 
Finally, the integrand needs to be averaged over every one of the $4!=24$ possible assignments between jets and partons.
This task is efficiently handled by a dedicated module that concentrates on the assignments dominating the average, as described in Sec.~\ref{sec:impl}.

\begin{figure}[!htbp]
  \centering
    \includegraphics[width=0.45\textwidth]{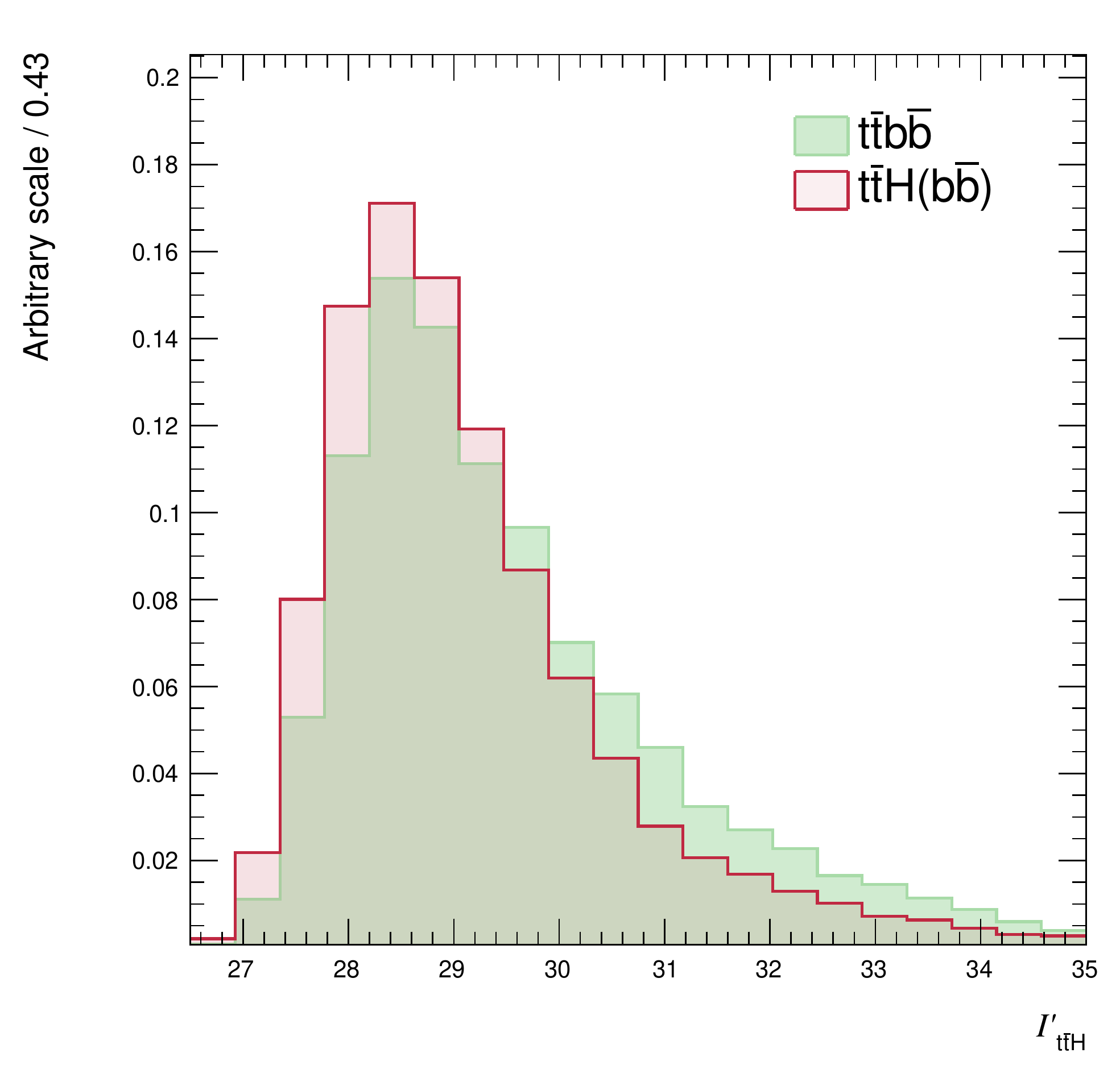}
    \includegraphics[width=0.45\textwidth]{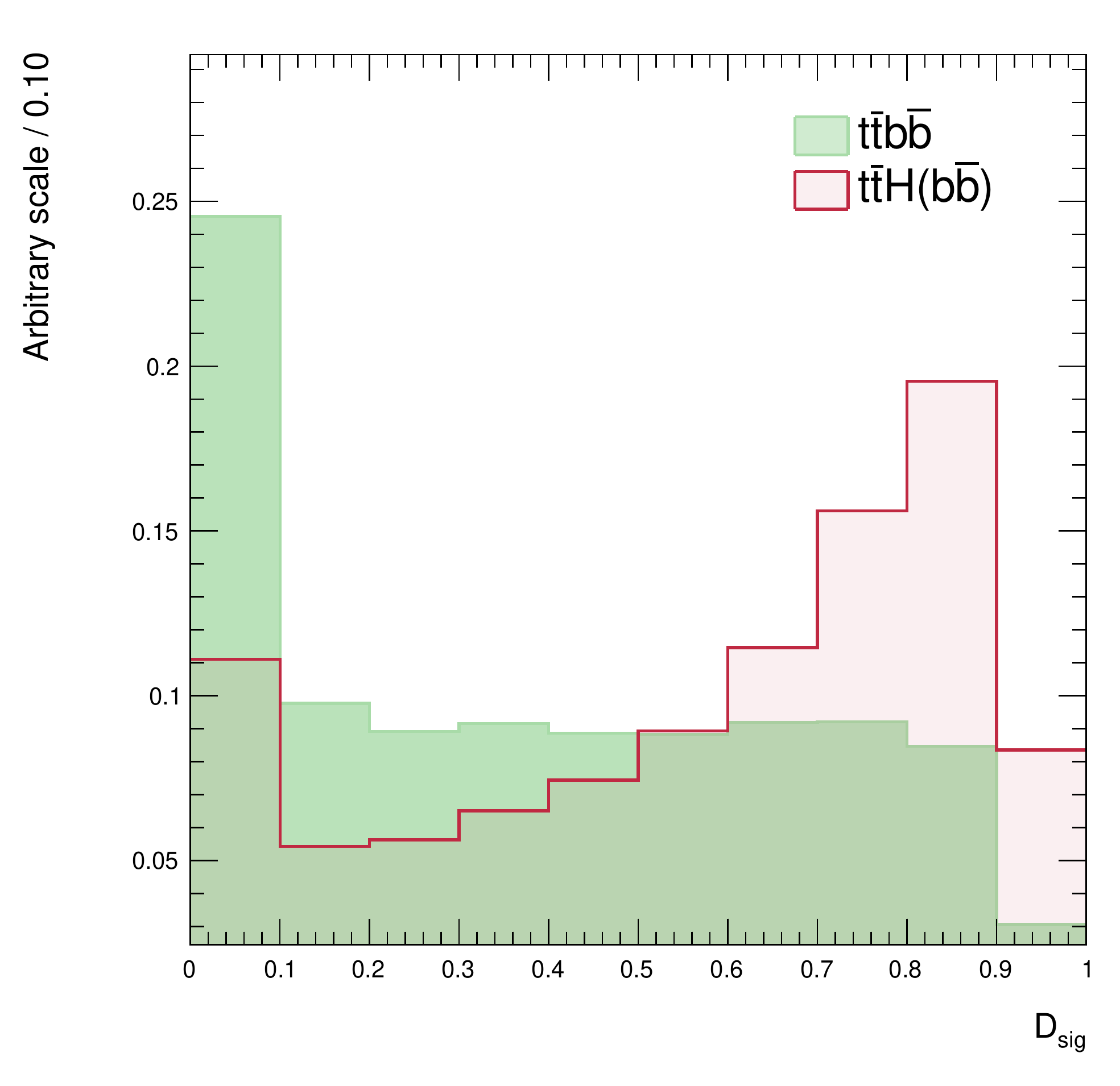}
    \caption{
    Left: signal information under the signal hypothesis ($\ttx\PH$). Right: discriminating variable built from the weights in the signal and background hypotheses. All distributions are normalised to unity.
    }
    \label{fig:tth}
\end{figure}

Figure~\ref{fig:tth} shows the normalised distributions of the event information $I'_{\ttx\PH}$
and a discriminating variable defined as 

\begin{equation}
    \label{eq:ttH} \mathcal{D}_{\text{sig}}(x) = \left( 1 + 
\frac{P(x|\ttx \Pqb\Paqb)}{P(x|\ttx\PH)} \right)^{-1}.
\end{equation}

By applying a requirement on $\mathcal{D}_{\text{sig}}$ such that 50\% of the $\ttx \PH$ signal is retained, 83\% of the $\ttx$ + jets background can be rejected.
As a comparison, using the invariant masses of pairs of b-tagged jets in the events, by choosing the pair of jets with mass closest to the true Higgs boson mass, would only reject 65\% of the background for the same signal efficiency.
The distributions shown in Fig.~\ref{fig:tth} (right) can be compared to those of Refs.~\cite{Khachatryan:2015ila,Sirunyan:2018mvw,Artoisenet:2013vfa}.

\section{Indicative performance figures}
\label{sec:perf}

We give approximate performance figures observed when computing weights for the different use cases presented above.
It should be clear that those numbers are indicative only, as the computation time strongly depends on the considered hypothesis and the parameters of the numerical integration algorithm.
Furthermore, these results were obtained using the functionalities available out-of-the-box in MoMEMta, and with matrix elements generated by our plugin for \textsc{MG5\_aMC@NLO}, which means no attempt whatsoever was made towards optimising the computation for these particular cases.
MoMEMta has been designed with the aim of being flexible, enabling users to implement simplifications or optimisations fit for their needs.
Note that tuning the parameters of the algorithms used for the numerical integration of the weights can have a strong impact on both the precision $p$ of the resulting integrals (which in turns impacts the power of the discriminant built from the weights), and the overall computation time $T$.
Generally, all other things being equal, the evaluation time $T$ scales roughly as $T \propto p^{-2}$.
The computation of the weights used in Sec.~\ref{sec:ttH} was carried out using two different integration algorithms available in the \textsc{Cuba} library: Vegas and Divonne.
The latter was found to yield substantially shorter completion times, without compromising the discrimination between signal and background with respect to the former.

In Tab.~\ref{tab:times} we give the average per-event computation times for the weights used in the examples of Sec.~\ref{sec:usecases}, along with the average relative precision on these weights reported by the integration algorithm.
These results were obtained on a computer cluster with an average per-core HS06 score\footnote{\url{https://w3.hepix.org/benchmarking.html}} of 9.1.
Table~\ref{tab:share} shows how much time is spent on the main elements of the computation.
These fractions are indicative and vary from event to event, but show that for complex hypotheses such as those considered in Sec.~\ref{sec:ttH}, the bottleneck in the computation is due to the evaluation of the matrix element. 

The memory consumption of MoMEMta is strongly linked to the way the integration algorithm is configured.
In practice, for the examples shown here, memory consumption was observed never to exceed 200 MB.

\begin{table}[!htpb]
    \caption{Average computation time of weights under the different hypotheses used in Sec.~\ref{sec:usecases}. 
    The average relative precision on the resulting weights are also given.}
    \label{tab:times}
\vspace*{\medskipamount}
\centering
\begin{tabular}{l c c } 
\hline \hline 
    Hypothesis & Avg. time & Avg. precision \\
\hline 
    $\Pg\Pg \to \rm{H} \to \PZ\PZ^* \to 4 \Pmu$, $\Pq\Paq \to \PZ\PZ/\PZ\Pgg^* \to 4 \Pmu$ & 0.6 ms & / \\
    $\Pqt\PWminus$, $\Paqt\PWplus$ & 4.5 s & 2.3\% \\
    $\ttx\PH(\Pqb\Paqb)$ (Vegas) & 140 s & 0.9\% \\
    $\ttx \Pqb\Paqb$ (Vegas) & 700 s & 1.0\% \\
    $\ttx\PH(\Pqb\Paqb)$ (Divonne) & 90 s & 0.5\% \\
    $\ttx \Pqb\Paqb$ (Divonne) & 600 s & 0.6\% \\
\hline \hline
\end{tabular}
\end{table}

\begin{table}[!htpb]
    \caption{Indicative shares of computation time due to the various elements entering the evaluation of weights under different hypotheses.
    The elements shown are the evaluation of the matrix element, the parton distribution functions, the transfer functions, and the generation of the phase space (including the changes of variables introduced in Sec.~\ref{sec:impl}).
    }
    \label{tab:share}
\vspace*{\medskipamount}
\centering
\begin{tabular}{l c c c c } 
\hline \hline 
    Hypothesis & Matrix element & PDF & Transfer functions & Phase space \\
\hline 
    $\Pqt\PWminus$, $\Paqt\PWplus$ & 28\% & 14\%  & 30\%  & 28\%  \\
    $\ttx\PH(\Pqb\Paqb)$           & 93\% & 1.3\% & 2.7\% & 2.6\% \\
    $\ttx \Pqb\Paqb$               & 98\% & 0.6\% & 0.6\% & 0.5\% \\
\hline \hline
\end{tabular}
\end{table}

\section{Summary}

We have presented MoMEMta, a modular software package to compute the
convolution integrals at the core of the MEM. Its modular structure
covers the needs of experimental analysis workflows at the LHC without
compromising the ease of use on simpler and smaller simulated samples
used for phenomenological studies.

The MEM has been used in HEP by both theoretical and experimental
communities with different purposes and levels of complexity ranging
from the evaluation of a matrix element on a reconstructed event to
the precise evaluation of model parameters through the use of the
properly normalised likelihood. We have described a few use cases of
the MEM for signal extraction in LHC analyses showcasing different
levels of complexity. From the most simple implementation, for Higgs
boson characterisation in the $\rm{H} \to \rm{ZZ}^* \to 4 \Pl$
channel, to complex final states such as $\ttx\PH$ production, MoMEMta
has proven to be sufficiently flexible to properly handle these different situations.

The main advantage of MoMEMta over past and existing tools comes from its 
modular design, that greatly improves on usability and flexibility.
MoMEMta is designed to offer a versatile and reusable framework for a wide range of applications of the MEM. 
While it is able to cover numerous use cases out of the box, the modular architecture of MoMEMta also enables users to easily extend its functionalities to handle situations we have not considered, while keeping the benefits of the I/O, 
configuration, and integration framework (a typical example would be the use of an optimised or simplified matrix element implementation).
As possible future developments, we are considering adding an interface to other matrix element libraries such as MCFM~\cite{MCFM} or Sherpa~\cite{Gleisberg:2008ta}, or enhancing the performance of the integration itself through the use of vector 
integrand with modified transfer function to evaluate the effect of systematic uncertainties, or through the use of 
machine-learning inspired integration algorithms~\cite{Bendavid:2017zhk}.

\section{Acknowledgments}

We warmly thank Andrea Giammanco and Olivier Mattelaer for their valued feedback.
This project is funded by FRS-FNRS (Belgian National Scientific Research Fund) IISN projects 4.4503.17 and 
4.4503.16.
MoMEMta is part of AMVA4NP, a project that has received funding from the European Horizon 2020 research and innovation 
programme under grant agreement \textnumero675440.
SW is supported through a FRIA grant by the F.R.S.-FNRS.
Computational resources have been provided by the supercomputing facilities of the Universit\'e catholique de Louvain (CISM/UCL) and the Consortium des \'Equipements de Calcul Intensif en F\'ed\'eration Wallonie Bruxelles (C\'ECI) funded by the Fond de la Recherche Scientifique de Belgique (F.R.S.-FNRS) under convention 2.5020.11.
This work would not have been possible without the help of the \textsc{MadWeight} team.
Special thanks to Matthias Komm who designed our logo.

This is a post-peer-review, pre-copyedit version of an article published in Eur. Phys. J. C. The final authenticated version is available online at: \url{http://dx.doi.org/10.1140/epjc/s10052-019-6635-5}.


\end{document}